
\magnification=1200
\hsize=31pc
\vsize=55 truepc
\hfuzz=2pt
\vfuzz=4pt
\pretolerance=5000
\tolerance=5000
\parskip=0pt plus 1pt
\parindent=16pt

\font\fourteenrm=cmr10 scaled \magstep2
\font\fourteeni=cmmi10 scaled \magstep2
\font\fourteenbf=cmbx10 scaled \magstep2
\font\fourteenit=cmti10 scaled \magstep2
\font\fourteensy=cmsy10 scaled \magstep2
\font\large=cmbx10 scaled \magstep1

\font\eightrm=cmr8
\font\eighti=cmmi8
\font\eightbf=cmbx8
\font\eightit=cmti8

\font\eightsy=cmsy8
\font\sixrm=cmr6
\font\sixi=cmmi6
\font\sixsy=cmsy6

\def\tenpoint{\def\rm{\fam0\tenrm}%
  \textfont0=\tenrm \scriptfont0=\sevenrm
                      \scriptscriptfont0=\fiverm
  \textfont1=\teni  \scriptfont1=\seveni
                      \scriptscriptfont1=\fivei
  \textfont2=\tensy \scriptfont2=\sevensy
                      \scriptscriptfont2=\fivesy
  \textfont3=\tenex   \scriptfont3=\tenex
                      \scriptscriptfont3=\tenex
  \textfont\itfam=\tenit  \def\it{\fam\itfam\tenit}%
  \textfont\slfam=\tensl  \def\sl{\fam\slfam\tensl}%
  \textfont\bffam=\tenbf  \scriptfont\bffam=\sevenbf
                            \scriptscriptfont\bffam=\fivebf
                            \def\bf{\fam\bffam\tenbf}%
  \normalbaselineskip=20 truept
  \setbox\strutbox=\hbox{\vrule height14pt depth6pt
width0pt}%
  \let\sc=\eightrm \normalbaselines\rm}
\def\eightpoint{\def\rm{\fam0\eightrm}%
  \textfont0=\eightrm \scriptfont0=\sixrm
                      \scriptscriptfont0=\fiverm
  \textfont1=\eighti  \scriptfont1=\sixi
                      \scriptscriptfont1=\fivei
  \textfont2=\eightsy \scriptfont2=\sixsy
                      \scriptscriptfont2=\fivesy
  \textfont3=\tenex   \scriptfont3=\tenex
                      \scriptscriptfont3=\tenex
  \textfont\itfam=\eightit  \def\it{\fam\itfam\eightit}%
  \textfont\bffam=\eightbf  \def\bf{\fam\bffam\eightbf}%
  \normalbaselineskip=16 truept
  \setbox\strutbox=\hbox{\vrule height11pt depth5pt width0pt}}
\def\fourteenpoint{\def\rm{\fam0\fourteenrm}%
  \textfont0=\fourteenrm \scriptfont0=\tenrm
                      \scriptscriptfont0=\eightrm
  \textfont1=\fourteeni  \scriptfont1=\teni
                      \scriptscriptfont1=\eighti
  \textfont2=\fourteensy \scriptfont2=\tensy
                      \scriptscriptfont2=\eightsy
  \textfont3=\tenex   \scriptfont3=\tenex
                      \scriptscriptfont3=\tenex
  \textfont\itfam=\fourteenit  \def\it{\fam\itfam\fourteenit}%
  \textfont\bffam=\fourteenbf  \scriptfont\bffam=\tenbf
                             \scriptscriptfont\bffam=\eightbf
                             \def\bf{\fam\bffam\fourteenbf}%
  \normalbaselineskip=24 truept
  \setbox\strutbox=\hbox{\vrule height17pt depth7pt width0pt}%
  \let\sc=\tenrm \normalbaselines\rm}

\def\today{\number\day\ \ifcase\month\or
  January\or February\or March\or April\or May\or June\or
  July\or August\or September\or October\or November\or
December\fi
  \space \number\year}

\newcount\secno      
\newcount\subno      
\newcount\subsubno   
\newcount\appno      
\newcount\tableno    
\newcount\figureno   

\normalbaselineskip=20 truept
\baselineskip=20 truept

\def\title#1
   {\vglue1truein
   {\baselineskip=24 truept
    \pretolerance=10000
    \raggedright
    \noindent \fourteenpoint\bf #1\par}
    \vskip1truein minus36pt}

\def\author#1
  {{\pretolerance=10000
    \raggedright
    \noindent {\large #1}\par}}

\def\address#1
   {\bigskip
    \noindent \rm #1\par}

\def\shorttitle#1
   {\vfill
    \noindent \rm Short title: {\sl #1}\par
    \medskip}

\def\jnl#1
   {\noindent \rm Submitted to: {\sl #1}\par
    \medskip}

\def\date
   {\noindent Date: \today\par
    \medskip}

\def\section#1
   {\vskip0pt plus.1\vsize\penalty-250
    \vskip0pt plus-.1\vsize\vskip24pt plus12pt minus6pt
    \subno=0 \subsubno=0
    \global\advance\secno by 1
    \noindent {\bf \the\secno. #1\par}
    \bigskip
    \noindent}

\def\subsection#1
   {\vskip-\lastskip
    \vskip24pt plus12pt minus6pt
    \bigbreak
    \global\advance\subno by 1
    \subsubno=0
    \noindent {\sl \the\secno.\the\subno. #1\par}
    \nobreak
    \medskip
    \noindent}

\def\subsubsection#1
   {\vskip-\lastskip
    \vskip20pt plus6pt minus6pt
    \bigbreak
    \global\advance\subsubno by 1
    \noindent {\sl \the\secno.\the\subno.\the\subsubno. #1}\null. }

\def\appendix#1
   {\vskip0pt plus.1\vsize\penalty-250
    \vskip0pt plus-.1\vsize\vskip24pt plus12pt minus6pt
    \subno=0
    \global\advance\appno by 1
    \noindent {\bf Appendix \the\appno. #1\par}
    \bigskip
    \noindent}

\def\subappendix#1
   {\vskip-\lastskip
    \vskip36pt plus12pt minus12pt
    \bigbreak
    \global\advance\subno by 1
    \noindent {\sl \the\appno.\the\subno. #1\par}
    \nobreak
    \medskip
    \noindent}

\def\figcaption#1
   {\global\advance\figureno by 1
    \noindent {\bf Figure \the\figureno.} \rm#1\par
    \bigskip}

\def\references
     {\vfill\eject
     {\noindent \bf References\par}
      \parindent=0pt
      \bigskip}

\def\refjl#1#2#3#4
   {\hangindent=16pt
    \hangafter=1
    \rm #1
   {\frenchspacing\sl #2
    \bf #3}
    #4\par}

\def\refbk#1#2#3
   {\hangindent=16pt
    \hangafter=1
    \rm #1
   {\frenchspacing\sl #2}
    #3\par}

 \def\numrefjl#1#2#3#4#5
   {\parindent=40pt
    \hang
    \noindent
    \rm {\hbox to 30truept{\hss #1\quad}}#2
   {\frenchspacing\sl #3\/
    \bf #4}
    #5\par\parindent=16pt}

\def\numrefbk#1#2#3#4
   {\parindent=40pt
    \hang
    \noindent
    \rm {\hbox to 30truept{\hss #1\quad}}#2
   {\frenchspacing\sl #3\/}
    #4\par\parindent=16pt}

\def\i{\ifmmode{\rm i}\else\char"10\fi}

\catcode`\@=11
\def\vfootnote#1{\insert\footins\bgroup
    \interlinepenalty=\interfootnotelinepenalty
    \splittopskip=\ht\strutbox 
    \splitmaxdepth=\dp\strutbox \floatingpenalty=20000
    \leftskip=0pt \rightskip=0pt \spaceskip=0pt \xspaceskip=0pt
    \noindent\eightpoint\rm #1\ \ignorespaces\footstrut\futurelet\next\fo@t}

\def\ind{\hbox to 5pc{}}
\def\eq(#1){\hfill\llap{(#1)}}

\def\deqn#1{\displ@y\halign{\hbox to \displaywidth
    {$\@lign\displaystyle##\hfil$}\crcr #1\crcr}}
\def\indeqn#1{\displ@y\halign{\hbox to \displaywidth
    {$\ind\@lign\displaystyle##\hfil$}\crcr #1\crcr}}
\def\indalign#1{\displ@y \tabskip=0pt
  \halign to\displaywidth{\ind$\@lign\displaystyle{##}$\tabskip=0pt
    &$\@lign\displaystyle{{}##}$\hfill\tabskip=\centering
    &\llap{$\@lign##$}\tabskip=0pt\crcr
    #1\crcr}}
\catcode`\@=12


\def\CMP{Commun. Math. Phys.}

\def\JSP{J. Stat. Phys.}

\def\NP{Nucl. Phys.}

\def\PR{Phys. Rev.}

\def\PRL{Phys. Rev. Lett.}

\def\ket#1{| #1 \rangle}
\def\aba{\vert a\vert}

\title {Matrix-product-groundstates for one-dimensional spin-1
quantum antiferromagnets\footnote*{Work performed
within the research program of the Sonderforschungsbereich\ 341,
K\"oln-Aachen-J\"ulich}}

\author{A.\ Kl\"umper, A.\ Schadschneider, and J.\ Zittartz}

\address{Institut f\"ur Theoretische Physik, Universit\"at zu K\"oln,
Z\"ulpicher Stra{\ss}e 77, D-50937 K\"oln, Federal
Republic of Germany}

\vskip2truecm\noindent
{\bf Abstract:} We have found the exact groundstate for a large class of
antiferromagnetic spin-1 models with nearest-neighbour interactions on a
linear chain. All groundstate properties can be calculated. The
groundstate is determined as a matrix product of individual site states
and has the properties of the Haldane scenario.

\vfill
\jnl{Europhys.\ Lett.}

\date
\vfill\eject
In 1983 Haldane [1] argued that there is a fundamental difference
between integer and half-integer spin chains (for a recent review, see [2]).
Certain quantum antiferromagnets with integral spin - including the isotropic
Heisenberg chain - should have the following properties:
\item{$(i)$} the groundstate is unique,
\item{$(ii)$} there is an energy gap between groundstate and excited states,
\item{$(iii)$} groundstate correlations have exponential decay.
\smallskip
This will be called 'Haldane scenario' in the following and is in contrast to
the behaviour of isotropic half-integral spin chains which are expected to
have no gap and algebraic decay of correlations. This scenario has been
verified in a number of experiments [3-6].
\smallskip
The first model for which all these properties could be proven rigorously was
introduced by Affleck, Kennedy, Lieb, and Tasaki [7]. This so-called VBS-model
(Valence Bond Solid) and some generalizations of it have subsequently been
studied intensively, see e.g.~[8-13] and references therein.

In this letter we present a large class of antiferromagnetic spin-1 chains for
which the groundstate can be given explicitly as a product of matrices. We like
to point out that our class differs from other models with product groundstates
as the well-known Majumdar-Gosh type models [14] (for further references, see
e.g.~[7,12]).

The most general spin-1 chain with nearest-neighbour interactions and the
following symmetries
\item{$(a)$} rotational invariance in the $x-y-$plane,
\item{$(b)$} invariance under $S^z\to -S^z$,
\item{$(c)$} translation and parity invariance, i.e.\ invariance under the
exchange $j \leftrightarrow j+1$,
\smallskip\noindent
can be written in the following form:
$$
\eqalign{
{\cal H}  &= \sum_{j=1}^L h_{j,j+1}, \cr
h_{j,j+1} &= \alpha_0 A_j^2+\alpha_1(A_jB_j+B_jA_j)+\alpha_2B_j^2+\alpha_3A_j
+\alpha_4B_j(1+B_j)\cr
&\quad +\alpha_5\left(\left(S_j^z\right)^2 + \left(S_{j+1}^z\right)^2\right)
+c, \cr}
\eqno(1)
$$
with real parameters $\alpha_j$, a constant $c$,  and periodic boundary
conditions. The nearest-neighbour interactions are
$$
\eqalign{
A_j &=S_j^xS_{j+1}^x+S_j^yS_{j+1}^y=S_j^+S_{j+1}^-+S_j^-S_{j+1}^+ \quad
\quad{\rm (transverse)}, \cr
B_j &= S_j^zS_{j+1}^z {\phantom{+S_j^yS_{j+1}^y=S_j^+S_{j+1}^-+S_j^-S_{j+1}^+
}}
\quad\quad {\rm (longitudinal)}.\cr}
\eqno(2)
$$
Here the raising and lowering operators are defined by $S_j^\pm={1\over
\sqrt{2}} \left(S_j^x\pm iS_j^y\right)$. Spin chains of the type (1) have been
studied extensively in recent years using various analytic and numerical
methods [15-22]. Only a few special cases of (1) are known to be exactly
solvable (for some references, see [7,12,13,19]).

In the following the constant $c$ in (1) will be adjusted to
have the groundstate eigenvalue of $h_{j,j+1}$ at 0. Thus we have
$$
h_{j,j+1}\geq 0 \ \Longrightarrow \  {\cal H} \geq 0,
\eqno(3)
$$
i.e.~all eigenvalues of ${\cal H}$ are non-negative.
\smallskip
We shall be interested in the antiferromagnetic case of the model where the
groundstate is characterized by $S_{total}^z=0$. In the following we show that
in a certain subspace of the $\alpha_j$-parameter space the antiferromagnetic
groundstate can be constructed in the form of a matrix product of single-site
states [13]. These states will be called {\bf M}atrix-{\bf P}roduct-{\bf
G}roundstates (MPG) in the following. In [13] we have shown that the
groundstate of the VBS-model and its $q$-deformed generalization [19] is a
special MPG.

Using the $S^z$ eigenstates $\ket{0}_j$ and $\ket{\pm}_j$ we define a local
$2\times 2$ matrix at each site $j$ by
$$
g_j=\left(\matrix{\ket{0}_j&-\sqrt{a} \ket{+}_j\cr
\sqrt{a} \ket{-}_j&-\sigma\ket{0}_j}\right)
\eqno(4)
$$
with nonvanishing parameters $a,\sigma \neq 0$ and the global
(antiferromagnetic) state
$$
\ket{\psi_0(a,\sigma)}={\rm Trace }\,  g_1\otimes g_2 \otimes ... \otimes g_L
\, .
\eqno(5)
$$

Here '$\otimes$' denotes usual matrix multiplication of 2$\times$2 matrices
with a tensor product of the matrix-elements. We shall show in an extended
paper [23] that the ansatz (4) is most general for the
antiferromagnetic region of (1) in which we are mainly interested here.
That $\ket{\psi_0}$ is antiferromagnetic, i.e. $S_{total}^z\ket{\psi_0}=0$,
can be checked easily.

We demand that the MPG $\ket{\psi_0(a,\sigma)}$ is groundstate of $\cal{H}$
with
eigenvalue 0, which requires certain conditions to be satisfied. Since we have
the  obvious implication
$$
h_{j,j+1}\ket{\psi_0(a,\sigma)}=0\ \Longrightarrow\
{\cal H}\ket{\psi_0(a,\sigma)}=0,
\eqno(6)
$$
and we have the product property (5) it is sufficient to show that
$$
h_{j,j+1}\left( g_j\otimes g_{j+1}\right)=0\, ,
\eqno(7)
$$
i.e.~each of the four elements of $g_j\otimes g_{j+1}$ which are
linearly independent is a groundstate of $h_{j,j+1}$. Calculating the
product in (7) explicitly one can show that (3),~(7) are satisfied provided
the following equalities
$$
\eqalign{
&1)\ \sigma={\rm sign}\, \alpha_3, \quad\qquad\quad \quad \quad
2)\ a\alpha_0=\alpha_3-\alpha_1, \cr
&3)\ \alpha_5=\vert\alpha_3\vert+\alpha_0
(1-a^2), \quad 4)\ \alpha_2=\alpha_0 a^2-2\vert\alpha_3\vert,\cr}
\eqno(8)
$$
and inequalities
$$
\eqalign{
&a\neq 0{\phantom{_4}},\quad \alpha_3\neq 0, \cr
&\alpha_4 > 0,\quad \alpha_0 >0 \cr}
\eqno(9)
$$
hold. The inequalities guarantee that all other eigenvalues of $h_{j,j+1}$
are strictly positive.  Thus $\ket{\psi_0(a,\sigma)}$ is groundstate of the
model (1) with eigenvalue 0 if only the restrictions (8) are satisfied. In
addition, if also the inequalities (9) are satisfied this groundstate is
unique for {\it any} chain length $L$ and there exists a finite gap
$\Delta>0$ to the excitations (in the thermodynamic limit $L\to \infty$) as
will be shown in detail in [23]. The first assertion can be understood from
the fact that the MPG (5) is an optimal state in the sense that it is the
product of all the local groundstates of $h_{j,j+1}$. The second assertion
can be proved by a slight generalization of the proof [7] for the VBS-model
which is recovered as the special case $a=2$, $\sigma=1$,
$\alpha_3=3\alpha_0>0$, $\alpha_2=-2\alpha_0$ and $\alpha_4=3\alpha_0$
[7,13]. Thus $(i)$ and $(ii)$ of the Haldane scenario are verified.

It is clear by continuity that on the boundaries (equality signs in the
inequalities (9)) $\ket{\psi_0(a,\sigma)}$ is still groundstate, but no
longer {\it unique}.

In order to verify point $(iii)$ of the Haldane scenario we calculate the
groundstate correlations explicitly. This can be done most easily using the
transfer-matrix method of [13] and is explained in detail in [23]. We find
for the longitudinal 2-site correlation
for $L\to \infty$  and $r\geq 2$
$$
\langle S_1^zS_r^z\rangle = -{a^2 \over (1-\aba)^2}\left({1-\aba \over 1+\aba}
\right)^r
\eqno(10)
$$
and for the transversal correlation
$$
\langle S_1^xS_r^x\rangle = -\aba \left[\sigma+{\rm sign}\, a\right]
\left({-\sigma \over 1+\aba}\right)^r.
\eqno(11)
$$
Furthermore we find
$$
\langle S_j^{x,y,z} \rangle = 0,\qquad
\langle \left(S_j^z\right)^2\rangle = {\aba \over 1+\aba}.
\eqno(12)
$$
Obviously the correlations (10), (11) decay exponentially with longitudinal
and transverse correlation lengths
$$
\xi_l^{-1}=\ln \Bigl| {1+|a|\over 1-|a|} \Bigr|, \quad
\xi_t^{-1}=\ln (1+|a|).
\eqno(13)
$$
The string-order-parameter [24-26] $\sigma^z =\lim_{r\to\infty}\left\langle -
S_1^z\exp\left(i\pi\sum_{j=2}^{r-1}S_j^z\right)S_r^z\right\rangle$ can also be
calculated easily. It is simply given by $\left\langle \left(S_j^z\right)^2
\right\rangle^2$ and thus $\sigma^z \ne 0$.

Now we briefly discuss the structure of the solution manifold ${\cal S}$
defined by (8) and (9) but refer to [23] for a detailed discussion. Apart
from the additive constant $c$ the model (1) is defined by six parameters
$\alpha_0,\ldots,\alpha_5$ including a trivial scale.
Two of the four conditions (8) are satisfied by fixing the two free parameters
$a,\sigma$ of the MPG (5), the remaining two conditions, i.e. 3),~4) of (8),
reduce the number of free parameters of the model to four. Thus ${\cal S}$
is a 4-dimensional parameter subspace. The parameters may be chosen as
$\alpha_0,\alpha_4>0$, $\alpha_3\neq 0$ and $\alpha_1$ such that $a\neq 0$.
Obviously ${\cal S}$ has external boundaries, $\alpha_0=0$ and $\alpha_4=0$,
and two internal boundaries, namely $\alpha_3=0$ and $a=0$.

At $\alpha_4=0$ the MPG is still a groundstate but it becomes degenerate
with the ferromagnetic eigenstates $\ket{+\cdots +}$ and $\ket{-\cdots -}$
which both have gapless spin-wave states adjacent to them (see also [23]).
We have a first-order phase transition to the ferromagnetic regime at
$\alpha_4=0$.

At $\alpha_0=0$ the parameter $a$ is not uniquely determined by (8) but
$\ket{\psi_0(a,\sigma)}$ spans a $L/2$-dimensional space of degenerate states.
One of these states is $\ket{0\cdots 0}$ which again has gapless spin-wave
states adjacent. Here we have a first-order transition to an
(antiferromagnetic)
regime.

Within ${\cal S}$ we have two internal boundaries $\alpha_3=0$ and $a=0$
where also phase transitions occur. At $\alpha_3=0$ $\sigma={\rm sign}
(\alpha_3)$ jumps from + to $-$ indicating a first-order transition. This
is seen explicitly in the behaviour of the transverse correlation $\langle
S_1^xS_r^x\rangle$ (11). The two MPGs $\ket{\psi_0(a,+)}$ and
$\ket{\psi_0(a,-)}$ are degenerate with each other and also with the states
$\ket{\pm 0\pm 0\cdots}$ and $\ket{0\pm 0\pm \cdots}$, indicating a vast
groundstate degeneracy.

At the other internal boundary $a=0$ the MPG becomes $\ket{\psi_0(0,\sigma)}
=\ket{0\cdots 0}$, i.e.\ all spins are in the $x-y$-plane.  As in the case
$\alpha_0=0$ we find low-lying spin-wave excitations and a vanishing gap.
As is obvious from (13), the correlation lengths diverge with $a\to 0$, thus
we have a critical transition.

In conclusion we note that we have found an interesting realistic model for
which the groundstate can be determined exactly and which shows a number of
different phases.

We also like to mention that a generalization of the above considerations
to arbitrary spin $S$ - including non-integral $S$ - is possible [23].
Also, using MPGs as variational states for the bilinear-biquadratic spin-1
chain (i.e.~the isotropic case of (1)) gives excellent results for the
groundstate energy [23].

\def\ZPhys{Z.\ Phys.\ }
\def\JPhys{J.\ Phys.\ }
\def\JSP{J.\ Stat.\ Phys.\ }
\def\CMP{Comm.\ Math.\ Phys.\ }
\def\PRL{Phys.\ Rev.\ Lett.\ }
\def\PR{Phys.\ Rev.\ }
\def\PhLett{Phys.\ Lett.\ }
\def\NP{Nucl.\ Phys.\ }
\def\Euro{Europhys.\ Lett.\ }
\references
\item{[1]} F.\ D.\ M.\ Haldane: \PhLett {\bf 93A}, 464 (1983); {\PRL}{\bf50},
1153  (1983); J.~Appl.~Phys.~{\bf 57}, 3359 (1985)
\item{[2]} I.~Affleck: \JPhys {\bf CM1}, 3047 (1989)
\item{[3]} J.~P.\ Renard, M.~Verdaguer, L.~P.~Regnault, W.~A.~C.~Erkelens,
J.~Rossat-Mignod, W.~G.~Stirling: \Euro {\bf 3}, 945 (1987)
\item{[4]} K.\ Katsumara, H.~Hori, T.~Takeuchi, M.~Date, M.~Yamagichi,
J.~P.~Renard: \PRL {\bf 63}, 86 (1989)
\item{[5]} K.\ Kakurai, M.~Steiner, R.~Pynn, J.~K.~Kjems: \JPhys {\bf CM3},
715 (1991)
\item{[6]} Z.~Tun, W.~J.~L.~Buyers, A.~Harrison, J.~A.~Rayne: \PR {\bf B43},
13331 (1991)
\item{[7]} I.\ Affleck, T.\ Kennedy, E.\ H.\ Lieb, H.\ Tasaki: {\PRL}{\bf 59},
799 (1987); {\CMP}{\bf 115}, 477 (1988)
\item{[8]} S.\ Knabe: \JSP {\bf 52}, 627 (1988)
\item{[9]} D.\ P.\ Arovas, A.\ Auerbach, F.\ D.\ M.\ Haldane: \PRL {\bf 60},
531 (1988)
\item{[10]} D.\ P.\ Arovas: \PhLett {\bf 137A}, 431 (1989)
\item{[11]} W.\ D.\ Freitag, E.\ M\"uller-Hartmann: {\ZPhys}{\bf B83}, 381
(1991); {\bf B88}, 279 (1992)
\item{[12]} M.\ Fannes, B\ Nachtergaele, R.\ F.\ Werner: \Euro {\bf 10}, 633
(1988); \CMP {\bf 144}, 443
(1992)
\item{[13]} A.\ Kl\"umper, A.\ Schadschneider, and J.\ Zittartz: {\JPhys}{\bf
A24}, L955 (1991); {\ZPhys}{\bf B87}, 281 (1992)
\item{[14]} C.~K.~Majumdar, D.~K.~Gosh: J.~Math.~Phys.~{\bf 10}, 1388 (1969)
\item{[15]} J.\ S\'olyom: \PR {\bf B36}, 8642 (1987)
\item{[16]} N.~Papanicolaou: \NP {\bf B305}, 367 (1988)
\item{[17]} G.~G\'omez-Santos: \PRL {\bf 63}, 790 (1989)
\item{[18]} K.~Chang, I.~Affleck, G.~W.~Hayden, Z.~G.~Soos: \JPhys {\bf CM1},
153 (1989)
\item{[19]} M.\ T.\ Batchelor, L.\ Mezincescu, R.\ I.\ Nepomechie,
V.\ Rittenberg: \JPhys {\bf A23}, L141 (1990)
\item{[20]} G.\ F\'ath, J.\ S\'olyom: \PR {\bf B44}, 11836 (1991); {\bf B47},
872 (1993)
\item{[21]} H.-J.\ Mikeska: \Euro {\bf 19}, 39 (1992)
\item{[22]} H.\ K\"ohler, R.\ Schilling: \JPhys {\bf CM4}, 7899 (1992)
\item{[23]} A.\ Kl\"umper, A.\ Schadschneider, J.\ Zittartz: (in
preparation)
\item{[24]} M.~den Nijs, K.~Rommelse: \PR {\bf B40}, 4709 (1989)
\item{[25]} S.~M.~Girvin, D.~P.~Arovas: Physica Scripta {\bf T27}, 156 (1989)
\item{[26]} T.\ Kennedy, H.\ Tasaki: \PR {\bf 45}, 304 (1992)
\vfill\eject
\end